\documentclass[final,1p,times]{elsarticle} 
\usepackage{graphicx}
\usepackage{epsfig}

\usepackage{amssymb}
\usepackage{setspace} 
\usepackage{lineno}

\journal{Nuclear Physics A}

\begin{document}

\begin{frontmatter}



\title{Femtoscopy of identified particles at STAR}


\author{ Neha Shah, for the STAR collaboration\fnref{label2}}
\fntext[label2]{A list of members of the STAR Collaboration and acknowledgements can be found at the end of this issue.}
\address{Department of Physics and Astronomy, UCLA, CA - 90095, USA}

\begin{abstract}

 We present new measurements of $\Lambda-\Lambda$  correlations, which are closely related to the $H$-dibaryon: a six quark state predicted by Jaffe~\cite{Jaffe}, in Au+Au collisions at $\sqrt{s_{NN}}$ = 200 GeV from the STAR experiment. Fits to different potential models are discussed and the extracted scattering length and effective range are also presented.
We also present a spherical harmonic decomposition of the pion-kaon correlation function for 0-5$\%$ centrality in Au+Au collisions at $\sqrt{s_{NN}} = 200$ GeV. Finally, we discuss results from azimuthal HBT for collision energies in the range 7.7-200 GeV.  

\end{abstract}

%\begin{keyword}



%\end{keyword}

\end{frontmatter}

%\linenumbers


%\section{Introduction}

Measurements of the correlation function for a pair of particles with small relative momenta can provide insight into the geometry and lifetime of the particle-emitting source in relativistic heavy ion collisions. The strong flow observed at RHIC suggests that the measured source size should be reduced relative to a source without flow. The source size also depends on the mass of a particle, with the heavier particles coming from earlier, hotter stages of the collision. If $\Lambda$s are produced in early stage of fireball expansion, they will be localized near the center of the system and the measured dimension of the system would be 2-3 fm. In addition to this, if the $H$-dibaryon is a resonance state, it would appear as a bump in the $\Lambda$-$\Lambda$ invariant mass spectra or if $H$-dibaryon is weakly bound state, it will lead to a depletion in the $\Lambda$-$\Lambda$ correlation near the threshold. Recent lattice QCD calculations from the HAL~\cite{HALQCD} and NPLQCD~\cite{HALQCD} collaborations indicate the existence of a bound $H$-dibaryon for a pion mass above the physical mass. The coalescence model predicts $H$-dibaryon production rate ranges from 10$^{-2}$ - 10$^{-4}$~\cite{dover} for the most central Au+Au collisions at $\sqrt{s_{NN}}=200$ GeV. 

The emission asymmetries between particles of different masses, which are closely related to the collective behavior of matter, can be studied by looking at non-identical particle correlations. The measurement of the non-identical particle correlation function in the three-dimensional $k^\ast$ space can reveal a space-time offset of one particle species with respect to another, which in our case are pions and kaons~\cite{mike}.

Hanbury-Brown-Twiss (HBT) interferometry allows us to determine the size of the pion-emitting source regions. The dependence of the HBT radius parameters on azimuthal angle relative to the reaction plane can be related to the freeze-out eccentricity. It is expected that the excitation function for freeze-out eccentricity will fall monotonically with increasing energy and any non-monotonic behaviour will indicate a change in the equation of state~\cite{excit}.
In this article we will mainly discuss recent measurements on the $\Lambda$-$\Lambda$ correlation function, the pion-kaon correlation function and azimuthal HBT. 

\section{$\Lambda$-$\Lambda$ correlation function } 

\begin{figure}[h]
\begin{center}
$\begin{array}{c@{\hspace{0.05in}}c}
\multicolumn{1}{l}{\mbox{}} &
\multicolumn{1}{l}{\mbox{}} \\ [-0.01cm]
\epsfxsize = 2.8in
\epsfysize=3.in
\epsffile{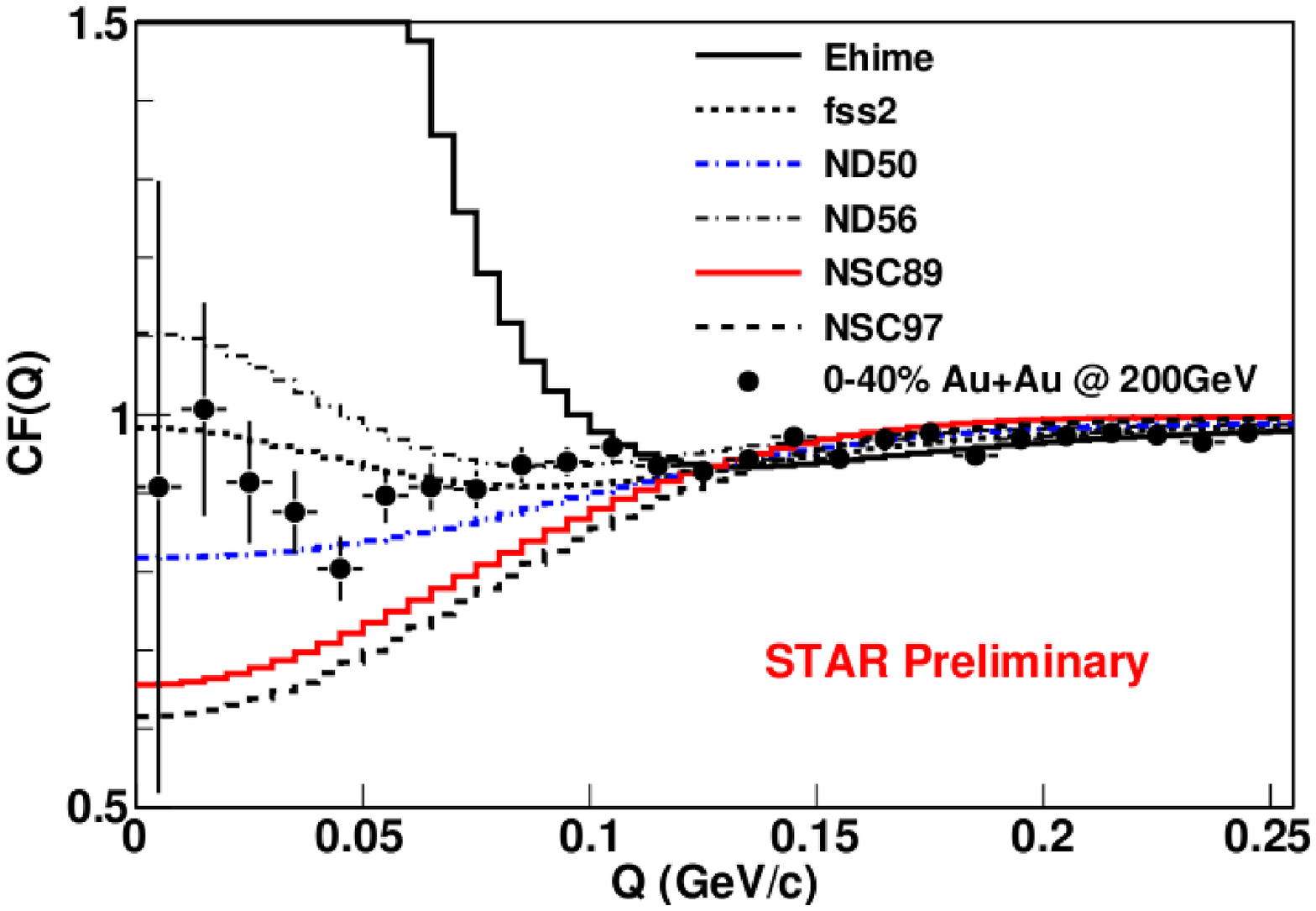} &
\epsfxsize=2.4in
\epsfysize=2.7in
\epsffile{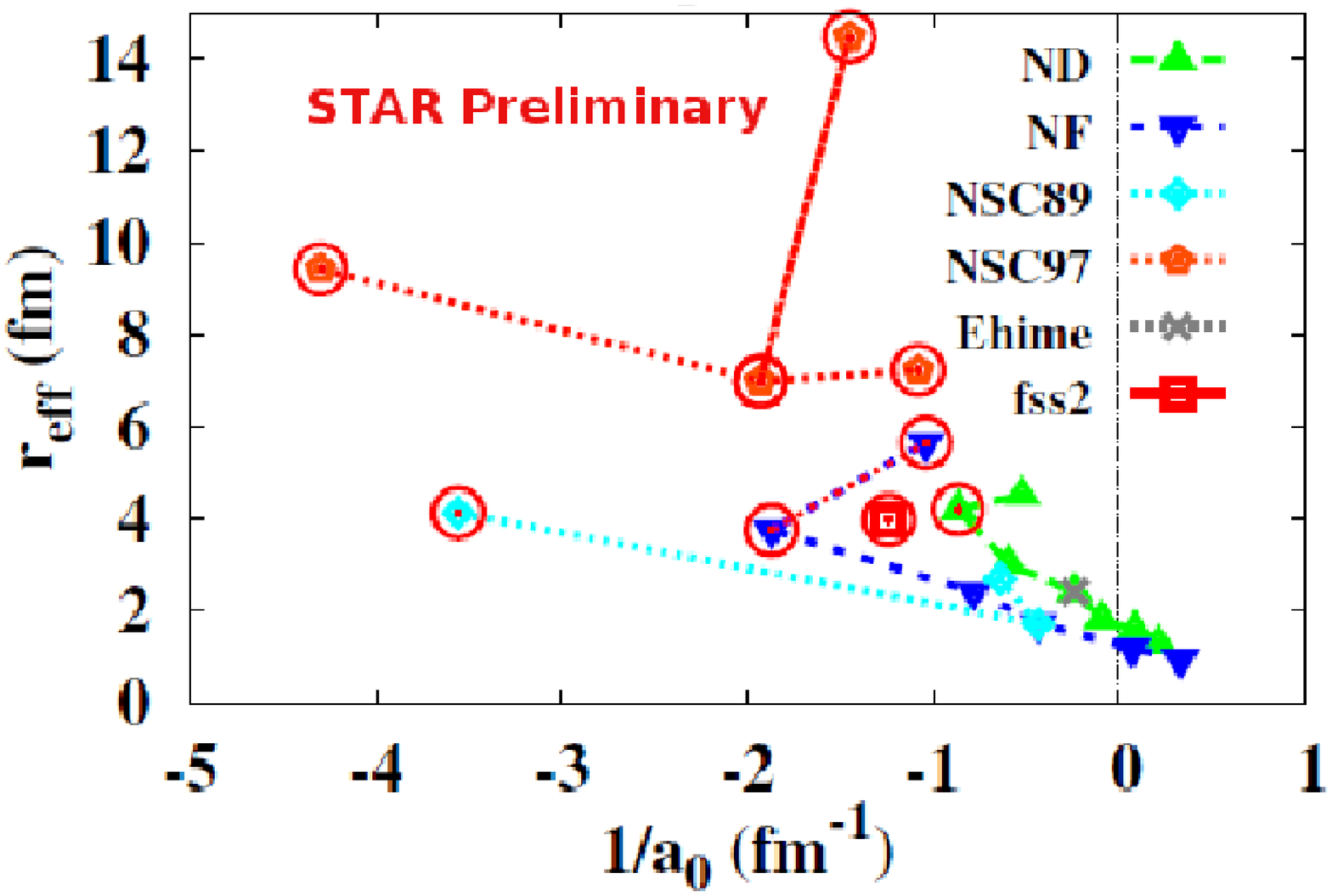} \\[0.1cm]
\mbox{(A)} & \mbox{(B)}
\end{array}$
\end{center}
\caption{(A) $\Lambda$-$\Lambda$ correlation function for 0-40$\%$ centrality in Au+Au collisions at $\sqrt{s_{NN}}= 200$ GeV. Curves correspond to fits done using different potential models  and (B) extracted scattering length and effective range (color online). }
\label{fig:LmLmCf}
\end{figure}

The measurement of $\Lambda$-$\Lambda$ correlations is important for hyperon-hyperon interaction studies as well as to pin down the existence of the $H$-dibaryon. Figure~\ref{fig:LmLmCf}(A) shows the inclusive $\Lambda$-$\Lambda$ correlation function for 0-40$\%$ centrality in Au+Au collisions at $\sqrt{s_{NN}}= 200$ GeV. The results have not been corrected for the residual correlations of$\Lambda$s from the $\Sigma$-baryon because of our limited understanding for hyperon-hyperon interaction potentials and the final state interactions. In Figure~\ref{fig:LmLmCf}(A), we  have also shown fits to the data with different potential models~\cite{Ohnishi}: Nijmegen Model D, F, Soft Core(89,97), Quark cluster model (fss2) and Ehime. The fss2 and Nijmegen Model D explains these data well. It is also important to note in fitting the data that residual correlations are taken into account in models. The $\Lambda$-$\Lambda$ interaction is attractive. The same models are used to extract the scattering length (a$_{0}$) and effective range (r$_{eff}$), which are shown in Figure~\ref{fig:LmLmCf}(B)~\cite{Ohnishi}. Most fit results give a negative scattering length, which may indicate non-existence of bound H-dibaryons.
\begin{figure}[h]
\epsfxsize=4.3in
\epsfysize=3.5in
\centerline{\epsfbox{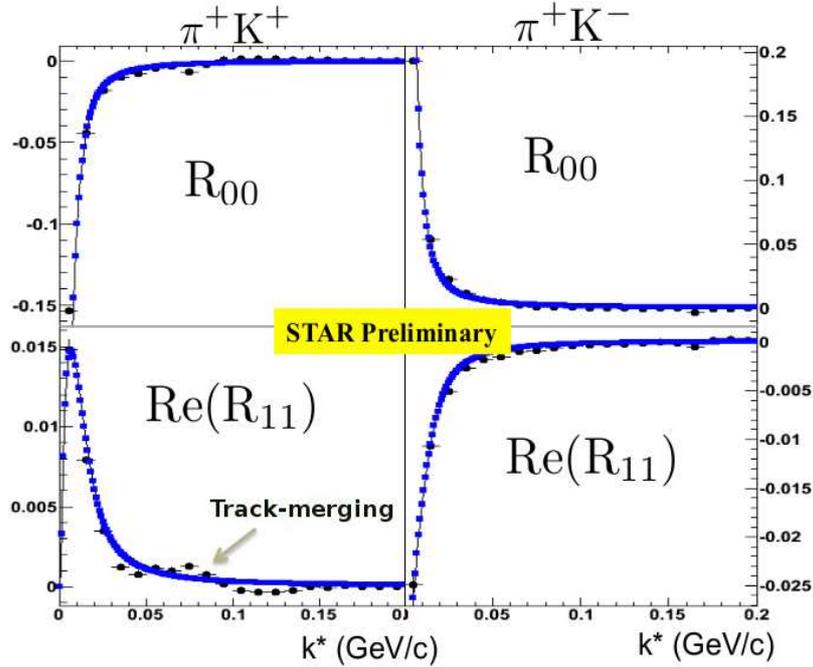}}
\caption{Spherical harmonic decomposition of pion-kaon correlation function for 0-5$\%$ centrality in Au+Au collision at $\sqrt{s_{NN}} = 200$ GeV. The solid line is the correlation function corresponding to a Gaussian separation distribution with a sigma of 14.1 fm and an average offset between the pions and kaons of 6 fm.}
\label{fig:pikcf}
\end{figure}
\section{Pion-Kaon correlations}

 Spherical harmonic decomposition of pion-kaon correlation~\cite{ref_shd} function for 0-5$\%$ centrality in Au+Au collisions at $\sqrt{s_{NN}} = 200$ GeV are shown in Figure~\ref{fig:pikcf}. The solid line in figure~\ref{fig:pikcf} is the correlation function corresponding to a Gaussian separation distribution with a sigma of 14.1 fm and an average offset between the pions and kaons of 6 fm. The top panel shows the 00 harmonic, which gives the size of the source and the bottom panel shows the 11 harmonic, which gives the separation between the average emission points of pions and kaons. The non-zero value of the 11 moment suggests that kaons are emitted farther out in the emission zone or at an earlier time. The observed offset between pions and kaons is roughly half of the source size, which is consistent with previous published measurements from Au+Au collisions at $\sqrt{s_{NN}} = 130$ GeV~\cite{SP130GeV}.

\section{Azimuthal HBT}

\begin{figure}[h]
\epsfxsize=4.3in
\epsfysize=2.9in
\centerline{\epsfbox{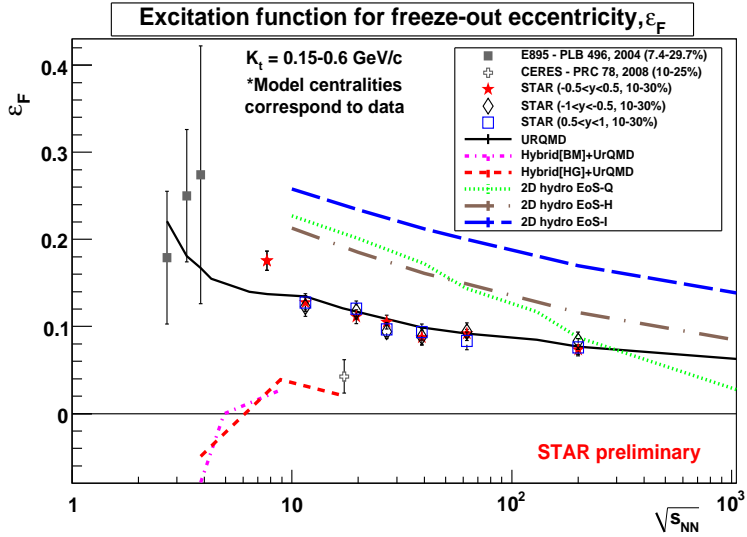}}
\caption{Freeze-out eccentricity, $\varepsilon_{f}$, as a function of $\sqrt{s_{NN}}$ for data and models~\cite{AZmodel} (color online).}
\label{fig:ExcFctn}
\end{figure}

Because of the initial eccentricity in the transverse plane in non-central heavy ion collisions, the participant zone forms an out-of-plane extended ellipsoid. At higher energies we expect stronger pressure gradients, which would cause the shape to become more spherical. Systems with a longer lifetime, as well, would achieve a more spherical shape or could even conceivably become extended in-plane. Based on these two considerations, we would expect the excitation function for the freeze-out eccentricity to fall monotonically with increasing energy. The excitation function of the freeze-out eccentricity for a centrality of 10-30$\%$ and rapidity ranges of [-0.5,0.5], [-1,-0.5] and [0.5,1] in Au+Au collisions as a function of  $\sqrt{s_{NN}}$ are given in Figure~\ref{fig:ExcFctn}. The STAR measurements are consistent 
with a monotonically decreasing trend. Comparison to models~\cite{AZmodel} show 
that the prediction from UrQMD comes closest to describing the measurements from STAR and the AGS. The hybrid models tend to come close to the CERES point but underpredict the rest, while 2D hydrodynamic predictions tend to overpredict the data at most energies. 

\section{Summary}

To summarize, the $\Lambda$-$\Lambda$ correlation function is presented. Fits to data with different potential models suggest that the $\Lambda$-$\Lambda$ interaction is attractive.  A negative scattering length gives indications towards the non-existence of bound H-dibaryons. A clear source asymmetry signal is observed in the pion-kaon correlation function and the offset is roughly half of the source size. The azimuthal HBT measurement shows a monotonic decrease for freeze-out eccentricity as a function of beam energy. 

%% References


%\bibliographystyle{model1b-num-names}

%% not want to use model1b-num-names.bst.

\noindent
\renewcommand{\bibfont}{\small}

\end{document}